\documentstyle[12pt]{article}
\def\ie{{\em i.e.}}
\def\eg{{\em e.g.}}

\def\bsg{B(b\to s\gamma)}

\def\beq{\begin{equation}}
\def\eeq{\end{equation}}

\catcode`\@=11 

\def\lsim{\mathrel{\mathpalette\@versim<}}
\def\gsim{\mathrel{\mathpalette\@versim>}}
\def\@versim#1#2{\vcenter{\offinterlineskip
    \ialign{$\m@th#1\hfil##\hfil$\crcr#2\crcr\sim\crcr } }}
\def\etal{{\em et. al.}}
\def\JL{J. L. Lopez}
\def\DVN{D. V. Nanopoulos}

\def\t1{{\tilde 1}}

\def\GeV{\,{\rm GeV}}

\def\to{\rightarrow}
\def\pb{\,{\rm pb}}
\def\fb{\,{\rm fb}}
\def\ipb{\,{\rm pb}^{-1}}
\def\NPB#1#2#3{Nucl. Phys. B {\bf#1} (19#2) #3}
\def\PLB#1#2#3{Phys. Lett. B {\bf#1} (19#2) #3}

\def\PRD#1#2#3{Phys. Rev. D {\bf#1} (19#2) #3}
\def\PRL#1#2#3{Phys. Rev. Lett. {\bf#1} (19#2) #3}
\def\PRT#1#2#3{Phys. Rep. {\bf#1} (19#2) #3}
\def\MODA#1#2#3{Mod. Phys. Lett. A {\bf#1} (19#2) #3}

\def\TAMU#1{Texas A \& M University preprint CTP-TAMU-#1}

\begin{document}
\begin{flushright}
\baselineskip=12pt
{CERN-TH.7107/93}\\
{CTP-TAMU-72/93}\\
{ACT-25/93}\\
{CERN-LAA/93-42}\\
\end{flushright}
%

\begin{center}
{\Large\bf First Constraints on SU(5)xU(1) Supergravity\\}
\vspace{0.2cm}
{\Large\bf from Trilepton Searches at the Tevatron\\}
\vglue 1cm
{JORGE L. LOPEZ$^{(a),(b)}$, D. V. NANOPOULOS$^{(a),(b),(c)}$,
GYE T. PARK$^{(a),(b)}$,\\}
{XU WANG$^{(a),(b)}$, and A. ZICHICHI$^{(d)}$\\}
\vglue 0.4cm
{\em $^{(a)}$Center for Theoretical Physics, Department of Physics, Texas A\&M
University\\}
{\em College Station, TX 77843--4242, USA\\}
{\em $^{(b)}$Astroparticle Physics Group, Houston Advanced Research Center
(HARC)\\}
{\em The Mitchel Campus, The Woodlands, TX 77381, USA\\}
{\em $^{(c)}$CERN, Theory Division, 1211 Geneva 23, Switzerland\\}
{\em $^{(d)}$CERN, 1211 Geneva 23, Switzerland\\}
\baselineskip=12pt

\vglue 0.5cm
{\tenrm ABSTRACT}
\end{center}
\vglue -0.2cm
{\rightskip=3pc
 \leftskip=3pc
\noindent
We present the first constraints on the parameter space of $SU(5)\times U(1)$
supergravity (in both no-scale and dilaton scenarios) which arise from the
recently announced limits on trilepton searches at the Tevatron. The trilepton
rate has been calculated for those points in parameter space which satisfy
not only the minimal theoretical and experimental LEP constraints, but also the
{\em combined} effect of the following indirect experimental constraints:
(i) the CLEO limits on the $b\to s\gamma$ rate, (ii) the long-standing limit on
the anomalous magnetic moment of the muon, (iii) the non-observation of
anomalous muon fluxes in underground detectors (``neutrino telescopes"), and
(iv) the electroweak LEP high-precision measurements in the form of the
$\epsilon_{1},\epsilon_b$ parameters. For $m_t=150\GeV$, the trilepton
constraint rules out some regions of parameter space with chargino masses as
high as $m_{\chi^\pm_1}\approx105\GeV$, although it is not possible to
establish a new absolute lower bound on the chargino mass. For $m_t=170\GeV$,
the simultaneous imposition of {\em all} of the above constraints excludes the
dilaton scenario completely, and leaves only a few allowed points in parameter
space in the no-scale scenario (with $m_{\tilde q}\approx m_{\tilde
g}\lsim285\GeV$). The five-fold increase in integrated luminosity expected in
the upcoming Tevatron run should probe some regions of parameter space with
chargino masses much beyond the reach of LEPII.}
\vspace{0.5cm}
\begin{flushleft}
\baselineskip=12pt
{CERN-TH.7107/93}\\
{CTP-TAMU-72/93}\\
{ACT-25/93}\\
{CERN-LAA/93-42}\\
November 1993
\end{flushleft}
\vfill\eject
\setcounter{page}{1}
\pagestyle{plain}

\baselineskip=14pt

\section{Introduction}
Searches for new physics at LEP have basically imposed a lower
bound of $\approx{1\over2}M_Z$ on the masses of all particles coupling to the
$Z$-boson with unsuppressed strength \cite{LEP}. Such is generally the case for
the sleptons, squarks, and charginos of low-energy supersymmetric models. No
further improvement in sensitivity is expected until LEPII turns on in early
1995. Supersymmetric particle searches have also been conducted at the
Tevatron, and until recently only in the strongly interacting sector (\ie,
squarks and gluino) \cite{Tevatron}. The weakly interacting sector had been
neglected because of the smaller production cross sections. This situation has
since changed because of the much increased integrated luminosity. In fact, the
prospects for supersymmetry detection through the trilepton signal
\cite{trileptons}, which occurs in the decays of neutralino-chargino pair
production, have been shown to be quite bright \cite{LNWZ} in $SU(5)\times
U(1)$ supergravity \cite{EriceDec92}. Moreover, because of the mass
correlations in this model, the direct search for charginos at the Tevatron
is a deeper probe of parameter space than the direct search for the heavier
squarks and gluino (\ie, $m_{\tilde q}\approx m_{\tilde g}\approx3.6
m_{\chi^\pm_1}$). Conversely, non-observation of charginos would entail strong
indirect lower bounds on the gluino and squark masses in this model.

It is worth noting that in the calculation of the trilepton signal, the
cross section $\sigma(p\bar p\to\chi^0_2\chi^\pm_1X)$ depends only on the
parameters of the neutralino-chargino sector (\ie, $M_2,\mu,\tan\beta$),
if the small contribution from the squark exchange diagrams is neglected.
However, the leptonic branching fractions of $\chi^0_2$ and $\chi^\pm_1$
depend additionally {\em and crucially} on the squark, slepton, and lightest
Higgs-boson masses. This proliferation of variables leads to a wealth of
possible outcomes, and within the Minimal Supersymmetric Standard Model(MSSM)
to a generic lack of predictability. In contrast, in $SU(5)\times U(1)$
supergravity  all model variables depend on only two parameters ($\tan\beta$
and $m_{\tilde g}$) and the top-quark mass, thus the predictions are quite
sharp and readily falsifiable.

In this paper we present the first constraints on the parameter space of
$SU(5)\times U(1)$ supergravity which arise from the recently announced limits
on trilepton searches at the Tevatron \cite{White,Kato,Kamon}. As predicted
\cite{LNWZ}, we show that this constraint is significant and for $m_t=150\GeV$
it rules out some regions of parameter space with chargino masses as high as
$m_{\chi^\pm_1}\approx105\GeV$, although it is not possible to establish a new
absolute lower bound on the chargino mass in $SU(5)\times U(1)$ supergravity.
The five-fold increase in integrated luminosity expected in the upcoming
Tevatron run should probe some regions of parameter space with chargino masses
much beyond the reach of LEPII.

Our calculations of the trilepton rate have been performed as described in
Ref.~\cite{LNWZ}, although without neglecting the $t$-channel (squark-exchange)
contribution to the cross section. More importantly, the parameter space
presently explored is much more constrained than in Ref.~\cite{LNWZ}, where
only the most basic theoretical/consistency, and experimental LEP constraints
were imposed (as described in detail in Ref.~\cite{aspects}). The present
parameter space has in addition been restricted by: (i) the CLEO limits on the
$b\to s\gamma$ rate \cite{bsgamma,bsg-eps}, (ii) the long-standing limit on the
anomalous magnetic moment of the muon \cite{g-2}, (iii) the non-observation of
anomalous muon fluxes in underground detectors (``neutrino telescopes")
\cite{NT}, and (iv) the electroweak LEP high-precision measurements in the form
of the $\epsilon_{1},\epsilon_b$ parameters \cite{ewcorr,bsg-eps,eps1-epsb}.
Furthermore, in the present analysis we consider two string-inspired universal
soft-supersymmetry-breaking scenarios for $SU(5)\times U(1)$ supergravity: the
no-scale \cite{LNZI} and dilaton \cite{LNZII} scenarios. Details of these
analyses and further experimental consequences will appear elsewhere
\cite{bigpaper}. An important consequence of the simultaneous imposition of all
the above constraints (trileptons included) is that for $m_t=170\GeV$, the
dilaton scenario is completely excluded, and only a few allowed points in
parameter space in the no-scale scenario remain (with $m_{\tilde q}\approx
m_{\tilde g}\lsim285\GeV$).

\section{SU(5)xU(1) Supergravity}
For practical purposes, the most important feature of $SU(5)\times U(1)$
supergravity is the minimality of parameters needed to describe the complete
low-energy supersymmetric spectrum and its interactions. The constraints
of supergravity and radiative electroweak symmetry breaking reduce the number
of parameters to four: the ratio of Higgs-boson vacuum expectation values
($\tan\beta$) and three universal soft-supersymmetry breaking parameters
($m_{1/2},m_0,A$) \cite{Erice93}. In addition, the top-quark mass ($m_t$)
plays an important role through the running of the mass parameters from the
unification scale down to the electroweak scale. Thus, until $m_t$ is measured
with some precision, it needs to be taken as a fifth parameter. In $SU(5)\times
U(1)$ supergravity we consider two string-inspired scenarios for the
{\em universal} soft-supersymmetry-breaking parameters, both of which belong to
the general no-scale supergravity framework \cite{LN}: (i) the no-scale
scenario \cite{Lahanas+EKNI+II}, where $m_0=A=0$, and (ii) the dilaton scenario
\cite{KL}, where $m_0={1\over\sqrt{3}}m_{1/2}$, $A=-m_{1/2}$. In this case, the
number of parameters is just two ($\tan\beta,m_{1/2}$) plus the top-quark mass.
For the latter we consider three values: $m_t=130,150,170\GeV$. In fact, the
present lower limit on the top-quark mass, obtained by combining the CDF and D0
lower bounds is 129 GeV \cite{Grannis}, and below we show that the case
$m_t=170\GeV$ is seriously constrained, if not completely excluded already.
Therefore, results of our calculations will be shown only for $m_t=150\GeV$.
The other values of $m_t$, as well as particular cases of the no-scale and
dilaton scenarios are considered in Ref.~\cite{bigpaper}. For $m_t=150\GeV$ we
find the following allowed range of $\tan\beta$: $2\lsim\tan\beta\lsim26\,(40)$
in the no-scale (dilaton) scenario.\footnote{The radiative breaking mechanism
requires $\tan\beta>1$, and the LEP lower bound on the lightest Higgs boson
mass ($m_h\gsim60\GeV$ \cite{LNPWZh}) is quite constraining for
$1<\tan\beta<2$.} The resulting parameter space for the no-scale \cite{LNZI}
and dilaton \cite{LNZII} scenarios consists of discrete pairs of points in the
$(\tan\beta,m_{1/2})$ plane. In practice, we trade the $m_{1/2}$ supersymmetric
mass scale parameter for the more readily measurable lightest chargino mass
$m_{\chi^\pm_1}$.

\begin{table}
\hrule
\caption{The approximate proportionality coefficients to the gluino mass, for
the various sparticle masses in the two supersymmetry breaking scenarios
considered. The $|\mu|$ coefficients apply for $m_t=150\GeV$ only.}
\label{Table1}
\begin{center}
\begin{tabular}{|c|c|c|}\hline
&no-scale&dilaton\\ \hline
$\tilde e_R,\tilde \mu_R$&$0.18$&$0.33$\\
$\tilde\nu$&$0.18-0.30$&$0.33-0.41$\\
$2\chi^0_1,\chi^0_2,\chi^\pm_1$&$0.28$&$0.28$\\
$\tilde e_L,\tilde \mu_L$&$0.30$&$0.41$\\
$\tilde q$&$0.97$&$1.01$\\
$\tilde g$&$1.00$&$1.00$\\
$|\mu|$&$0.5-0.7$&$0.6-0.8$\\\hline
\end{tabular}
\end{center}
\hrule
\end{table}

In the scenarios we consider all sparticle masses scale with the gluino mass,
with a mild $\tan\beta$ dependence (except for the third-generation squark and
slepton masses). In Table~\ref{Table1} we give the approximate
proportionality coefficient (to the gluino mass) for each sparticle mass. Note
that the relation $2m_{\chi^0_1}\approx m_{\chi^0_2}\approx m_{\chi^\pm_1}$
holds to good approximation. The third-generation squark and slepton masses
also scale with $m_{\tilde g}$, but the relationships are smeared by a strong
$\tan\beta$ dependence, and are therefore not shown in Table~\ref{Table1}.
{}From this table one can (approximately) translate any bounds on a given
sparticle mass on bounds on all the other sparticle masses.

\section{Constraints on the parameter space}
The parameter space for $SU(5)\times U(1)$ supergravity in the no-scale and
dilaton scenarios has been obtained in Refs.~\cite{LNZI,LNZII}. However, these
allowed points satisfy only the theoretical and consistency constraints, plus
the most basic experimental constraints from new particle searches at LEP
\cite{aspects}. Several indirect experimental constraints have been recently
considered in the context of $SU(5)\times U(1)$ supergravity, although until
now not in a ``global" way, \ie, not all constraints applied at once. These
constraints are significant and are now described in turn.
\begin{description}
\item (i) $b\to s\gamma$ : This rare radiative flavor-changing-neutral-current
(FCNC) decay has been observed by the CLEOII Collaboration in the following
95\% CL allowed range \cite{Thorndike}
\beq
\bsg=(0.6-5.4)\times10^{-4}.\label{bsg}
\eeq
The predictions for $\bsg$ in the no-scale scenario were given in
Ref.~\cite{bsgamma} and for the dilaton scenario in Ref.~\cite{bsg-eps}.
The experimental bound in Eq.~(\ref{bsg}) was seen to be quite restrictive
because the model predictions could be well above the Standard Model values,
as well as much suppressed relative to the SM case. For the case of
$m_t=150\GeV$, the points in parameter space excluded by this constraint
are represented by plus signs ($+$) in Fig.~1a (2a) for the no-scale (dilaton)
scenario. In both scenarios there are points (mostly for
$m_{\chi^\pm_1}<100\GeV$) where this constraint overlaps with the $(g-2)_\mu$
constraint considered next, and the resulting symbol is the overlap of a
plus sign ($+$) with a cross sign ($\times$). The case of $m_t=170\GeV$ is
shown in Figs.~1b,2b and entails similar constraints on the parameter space,
although these are harder to appreciate in the figures because of the
overwhelming effect of the $\epsilon_{1,b}$ constraints.
\item (ii) $(g-2)_\mu$ : The supersymmetric contributions to the anomalous
magnetic moment of the muon in $SU(5)\times U(1)$ supergravity have been
obtained in Ref.~\cite{g-2}. Comparing the long-standing experimental value
for $(g-2)_\mu$ with the most accurate determination of the Standard Model
contribution, one can determine a 95\%CL allowed interval for the
supersymmetric contribution. The latter contribution is greatly enhanced for
large values of $\tan\beta$ (which become constrained) and can easily exceed
the electroweak contribution and the hadronic uncertainties within the SM.
The points in parameter space excluded by this constraint are represented by
cross signs ($\times$) in Figs.~1,2. As for the $\bsg$ constraint, the
$(g-2)_\mu$ constraint is also harder to appreciate in the $m_t=170\GeV$ case
(\ie, in Figs.~1b,2b).
\item (iii) ``Neutrino telescopes": If neutralinos (the lightest supersymmetric
particle, which is stable) are present in the galactic halo in significant
numbers, as required to solve the halo dark matter puzzle \cite{Turner}, then
some would be captured by the Sun and Earth cores. Subsequent neutralino
annihilations will produce a shower of decay particles, but only the
high-energy neutrinos could escape the core vicinity. These neutrinos could
then interact with rock beneath underground detectors (called neutrino
telescopes) and produce upwardly-moving muon events in the detector. The
computation of these muon fluxes in $SU(5)\times U(1)$ supergravity has been
carried out in Ref.~\cite{NT}. The experimental limits on muon fluxes above
background are not very restrictive at the moment, but nonetheless the points
in Figs.~1,2 denoted by diamond symbols ($\diamond$) are excluded at the
90\%CL. These excluded points occur mostly for the dilaton scenario and for
$m_{\chi^\pm_1}\approx100\GeV$ (where neutralino capture by the Earth is
enhanced by the $^{56}{\rm Fe}$ nucleus \cite{NT}).
\item (iv) $\epsilon_1,\epsilon_b$ parameters : Precision electroweak
measurements at LEP can be expressed in terms of four parameters
($\epsilon_{1,2,3,b}$ \cite{ABCetc}) which are calculated from various one-loop
diagrams. The supersymmetric contributions to these parameters have been
calculated in $SU(5)\times U(1)$ supergravity and found to be constraining only
for $\epsilon_1$ \cite{ewcorr,bsg-eps} and $\epsilon_b$ \cite{eps1-epsb}. Both
of these parameters depend quadratically on $m_t$ and at the 90\%CL only values
of $m_t\gsim160\GeV$ are constrained experimentally \cite{eps1-epsb}.
In fact, Figs.~1b,2b show that in order for the case $m_t=170\GeV$ to give
values of $\epsilon_{1,b}$ within the experimental limits, it is required that
the chargino be light ($m_{\chi^\pm_1}\lsim70\GeV$) and $\tan\beta\gsim4$.
\end{description}

\section{The trilepton signal}
It had been suggested that searches for weakly interacting supersymmetric
particles at the Tevatron (and hadron colliders in general) could be a worthy
pursuit \cite{trileptons}. This expectation was investigated in the no-scale
scenario of $SU(5)\times U(1)$ supergravity in Ref.~\cite{LNWZ} and found to
be exceptionally well suited for probing the parameter space of this model.
The process of interest is $p\bar p\to \chi^0_2\chi^\pm_1 X$, where both
neutralino and chargino decay leptonically: $\chi^0_2\to\chi^0_1 l^+l^-$, and
$\chi^\pm_1\to \chi^0_1 l^\pm\nu_l$, with $l=e,\mu$. In Ref.~\cite{LNWZ}, the
$t$-channel (squark-exchange) contribution to $\sigma(p\bar p\to
\chi^0_2\chi^\pm_1 X)$ was neglected. Here we include this contribution, which
is found to be small (at most a $\pm10\%$ effect) and decreasing with
increasing squark masses. The irreducible backgrounds for this process are very
small, the dominant one being $p\bar p\to W^\pm Z\to
(l^\pm\nu_l)(\tau^+\tau^-)$ with a cross section into trileptons of
$(\sim1\pb)({2\over9})(0.033)(0.34)^2\sim1\fb$. Much larger ``instrumental"
backgrounds exist when for example in $p\bar p\to Z\gamma$, the photon
``converts" and fakes a lepton in the detector; with the present sensitivity,
suitable cuts have been designed to reduce this background to acceptable levels
\cite{White}.

The trilepton signal is larger in the no-scale scenario because the charged
sleptons which mediate some of the decay channels can be on-shell and the
leptonic branching ratios are significantly enhanced (as large as $2\over3$)
relative to a situation with heavier sparticles in the dilaton scenario, where
the $W,Z$-exchange channels tend to dominate and the leptonic branching
fractions are smaller \cite{LNWZ}. The results of these calculations for the
case of $m_t=150\GeV$ are shown in Fig.~3 for both scenarios. Note that for
light chargino masses, the trilepton signal can be rather small in the no-scale
model. This occurs when the neutralino leptonic branching fraction is
suppressed because the sneutrinos are on-shell and the
$\chi^0_2\to\nu\tilde\nu$ channel dominates.

In Fig.~3 we have also shown the very recent limits obtained by the D0
\cite{White} and CDF \cite{Kato,Kamon} Collaborations. Both these sets of
data are preliminary and the D0 limit is weakened because of the present
inability to remove one candidate event from the data sample, and because the
full data set has not yet been analyzed. The curves become more restrictive for
larger values of $m_{\chi^\pm_1}$ because the efficiency for detecting
charginos and neutralinos increases with their masses. The integrated
luminosity corresponds roughly to $15\ipb$ for D0 and $18\ipb$ for CDF.

Clearly, at least for $\mu<0$ (and $m_t=150\GeV$), the Tevatron (CDF) data is
restrictive.  Most interestingly, there are excluded points in parameter space
of the no-scale scenario (with $\mu<0$) with chargino masses as high as
$m_{\chi^\pm_1}\approx105\GeV$. These points are already beyond the reach of
LEPII. The points in parameter space excluded by the trilepton constraint are
shown in Figs.~1,2 as fancy star symbols. By construction, these points are not
excluded by any of the previously discussed constraints. Because of the
sometimes suppressed leptonic branching fractions, in general it is not
possible to obtain a new absolute lower bound on the chargino mass: for
$m_t=150\GeV$, combining all constraints we obtain
$m_{\chi^\pm_1}\gsim65\,(45)\GeV$ and $m_{\chi^\pm_1}\gsim50\,(60)\GeV$ in the
no-scale and dilaton scenarios for $\mu>0\,(\mu<0)$ respectively.

In the case of $m_t=170\GeV$, the imposition of the {\em combined} constraints
discussed above is so strong that the very few points in parameter space
remain allowed. In fact, in the dilaton scenario (see Fig. 2b) the remaining
points (with $m_{\tilde q}\approx m_{\tilde g}\approx195-220\GeV$) can be
excluded by the CDF lower bound on the squark and gluino masses
($m_{\tilde q}^{exp}\approx m_{\tilde g}^{exp}\gsim220\GeV$ \cite{Tevatron}).
In the no-scale scenario, for $m_t=170\GeV$ (see Fig.~1b) one has
$m_{\chi^\pm_1}\lsim70\,(60)\GeV$ and $m_{\tilde
q}\approx m_{\tilde g}\lsim260\,(285)\GeV$ for $\mu>0\,(\mu<0)$, and the
analysis of the 92--93 CDF and D0 data (once completed) could possibly exclude
all these points altogether.

\section{Conclusions and outlook}
We have calculated the effect of several experimental constraints on the
parameter space of $SU(5)\times U(1)$ supergravity. An important result is
that the case $m_t=170\GeV$ is allowed only for very light chargino masses
($m_{\chi^\pm_1}\lsim70\GeV$) and $m_{\tilde q}\approx m_{\tilde
g}\lsim285\GeV$ in the no-scale scenario, and is completely excluded in the
dilaton scenario. This is a consequence of the combined effect of {\em all}
constraints. Of the indirect constraints, probably the future increase in
sensitivity of $\bsg$ at CLEO is the most effective way to explore the
parameter space in the near future. Furthermore, starting in late 1994, the new
Brookhaven E821 experiment should be able to eventually constrain the parameter
space decisively with the high-precision measurement of $(g-2)_\mu$.

The direct trilepton search at the Tevatron looks quite promising as well. With
the five-fold increase in statistics during 1994, the reach in chargino masses
is limited by the handling of the backgrounds. In principle, if the backgrounds
could be suppressed to levels below the signal, the experimental limits shown
on Fig.~3 could drop by a factor of 5, entailing a probe of the parameter
space with chargino masses as high as $m_{\chi^\pm_1}\approx150\GeV$. Of
course, because of the details of the model, some points in parameter space
with lighter chargino masses would remain unexplored. However, what matters is
the fraction of the parameter space which is explored, and not which portion of
it one is able to explore first. The remaining lighter chargino regions
($m_{\chi^\pm_1}<100\GeV$) will be fully covered at LEPII. Squark and gluino
searches at the Tevatron may also become relevant with the increase in
integrated luminosity in the coming run. For $m_t=150\GeV$, the present
experimental limits \cite{Tevatron} do not constrain the parameter space of
$SU(5)\times U(1)$ supergravity in any significant way. However, for the
$m_t=170\GeV$ case the present squark and gluino mass limits supplement the
previously discussed constraints such as to exclude the dilaton scenario
altogether.

\section*{Acknowledgements}
We would like to thank James White and Teruki Kamon for very useful discussions
and for providing us with the latest experimental data.
This work has been supported in part by DOE grant DE-FG05-91-ER-40633. The work
of G.P. and X. W. has been supported by a World Laboratory Fellowship.

\newpage

\noindent{\large\bf Figure Captions}
\begin{description}
\item Figure 1: The parameter space of the no-scale scenario in $SU(5)\times
U(1)$ supergravity in the $(m_{\chi^\pm_1},\tan\beta)$ plane for (a)
$m_t=150\GeV$ and (b) $m_t=170\GeV$. The periods indicate points that passed
all constraints, the pluses fail the $\bsg$ constraint, the crosses fail the
$(g-2)_\mu$ constraint, the diamonds fail the neutrino telescopes (NT)
constraint, the squares fail the $\epsilon_1$ constraint, the fancy pluses
fail the $\epsilon_b$ constraint, and the fancy stars fail the
trilepton constraint. The dashed line indicates the direct reach of LEPII for
chargino masses. Note that when various symbols overlap a more complex symbol
is obtained.
\item Figure 2: The parameter space of the dilaton scenario in $SU(5)\times
U(1)$ supergravity in the $(m_{\chi^\pm_1},\tan\beta)$ plane for (a)
$m_t=150\GeV$ and (b) $m_t=170\GeV$. The periods indicate points that passed
all constraints, the pluses fail the $\bsg$ constraint, the crosses fail the
$(g-2)_\mu$ constraint, the diamonds fail the neutrino telescopes (NT)
constraint, the squares fail the $\epsilon_1$ constraint, the fancy pluses
fail the $\epsilon_b$ constraint, and the fancy stars fail the
trilepton constraint. The dashed line indicates the direct reach of LEPII for
chargino masses. Note that when various symbols overlap a more complex symbol
is obtained.
\item Figure 3: The trilepton cross section at the Tevatron ($p\bar p\to
\chi^0_2\chi^\pm_1 X$; $\chi^0_2\to\chi^0_1 l^+l^-$, $\chi^\pm_1\to \chi^0_1
l^\pm\nu_l$, with $l=e,\mu$) in $SU(5)\times U(1)$ supergravity for both
no-scale and dilaton scenarios. The CDF and D0 95\%CL limits are shown. Note
that some points with $m_{\chi^\pm_1}>100\GeV$ (in the no-scale case for
$\mu<0$) are excluded.
\end{description}

\end{document}